# A response to: "NIST experts urge caution in use of courtroom evidence presentation method"

*Geoffrey Stewart Morrison*

Reader in Forensic Speech Science, Centre for Forensic Linguistics, Aston University

Version 2017-10-16a

A press release from the National Institute of Standards and Technology (NIST) could potentially impede progress toward improving the analysis of forensic evidence and the presentation of forensic analysis results in courts in the United States and around the world. "NIST experts urge caution in use of courtroom evidence presentation method" [1] was released on October 12, 2017, and was picked up by the phys.org news service.[2] It argues that, except in exceptional cases, the results of forensic analyses should not be reported as "likelihood ratios". The press release, and the journal article by NIST researchers Steven P. Lund & Harri Iyer on which it is based,[3] identifies some legitimate points of concern, but makes a strawman argument and reaches an unjustified conclusion that throws the baby out with the bathwater.

Properly understood, the likelihood ratio framework describes what is logically necessary for a forensic scientist to evaluate the strength of forensic evidence. Imagine that eyewitnesses to a crime say that the offender had blond hair and a suspect is arrested who also has blond hair (and let us imagine a simplified world in which eyewitnesses are not mistaken, blond is clearly distinct from other hair colors, and people do not change the color of their hair or wear wigs). The forensic scientist is asked to evaluate the strength of the hair color evidence and only that evidence (the jury gets to consider all the evidence, but each piece of evidence should be independently analyzed by a different forensic scientist). Does the fact that both the offender and the suspect have blond hair mean that the offender is the suspect? Of course not! The suspect and offender are very similar with respect to hair color, but the forensic scientist also has to consider how typical blond hair is. The forensic scientist has to assess not only the probability that the offender would have blond hair if they were the suspect, but also the probability that the offender would have blond hair if they were instead someone selected at random from the relevant population. The first probability divided by the second probability is known as the "likelihood ratio". If the crime had been committed in Stockholm and the population of Stockholm was treated as the relevant population, the value of the likelihood ratio would be very different than if the crime had been committed in Beijing and the population of Beijing was treated as the relevant population. This is the essential logic of the likelihood ratio framework, the forensic scientist has to probabilistically assess both similarity and typicality. It doesn't matter whether the term "likelihood ratio" is used, this is logically what must be done. Knowing that both the offender and suspect have blond hair is meaningless unless one knows how common blond hair is in the relevant population. The 2016 report "Forensic science in criminal courts: Ensuring scientific validity of feature-comparison methods" by President Obama's Council of Advisors on Science and

---

[1] https://www.nist.gov/news-events/news/2017/10/nist-experts-urge-caution-use-courtroom-evidence-presentation-method

[2] https://phys.org/news/2017-10-nist-urges-caution-courtroom-evidence.html

[3] Lund S.P., Iyer H. (2017). Likelihood ratio as weight of forensic evidence: A closer look. *Journal of Research of National Institute of Standards and Technology*, 122, Article 27. https://doi.org/10.6028/jres.122.027

Technology (PCAST)[4] was highly critical of current practice in several branches of forensic science. Although it did not call it by name, one of the report's major recommendations was the adoption of the likelihood ratio framework (see "A comment on the PCAST report: Skip the 'match'/'non-match' stage" authored by myself and eighteen others).[5]

A practical problem is deciding what constitutes the relevant population in a particular case. Imagine that the jury believed that the relevant population was the population of Stockholm, but the forensic scientist believed that it was the population of Beijing. If the forensic scientist calculated typicality with respect to the population of Beijing, but the jury believed the calculation was with respect to the population of Stockholm, then the forensic scientist's statement as to the strength of the evidence would be highly misleading to the jury. In order to avoid such miscommunication, the forensic scientist must clearly explain to the jury what the forensic scientist has adopted as the relevant population. The jury can then (1) decide if the forensic scientist's choice is appropriate, and (2) understand the meaning of the likelihood ratio value presented by the forensic scientist.

Can mismatches in forensic scientists' and juries' assumptions cause communication problems and misunderstandings? Certainly they can and do. But this is not a problem which only affects the likelihood ratio framework, and, in contrast to what Lund & Iyer suggest, it is not a reason to reject the likelihood ratio framework. There are a range of difficulties in communicating the meaning of forensic likelihood ratios to juries. The solution is not to reject the use of the likelihood ratio framework, but to conduct more research on ways to improve understanding. Should a forensic scientist present something that is easy to understand but incorrect? An expert witness who has sworn to tell the truth cannot present something that they know to be incorrect. They must present what is correct even if it is challenging to communicate.

Another practical problem is assessing the degree of similarity and the degree of typicality. The forensic scientist will use statistical models to calculate the probabilities associated with both similarity and typicality. Let us focus on typicality: The forensic scientist looks at how many people in a relatively small group of people have blond hair and extrapolates that to estimate the proportion of the entire population who have blond hair. The small group is known as a sample of the population. If the size of the sample is too small, or the people in that sample are actually not representative of the population as a whole, then the estimated proportion of people in the population who have blond hair could be far from the true proportion. Another forensic scientist who uses a different sample of the relevant population could get a substantially different answer. The only way to know the true answer would be to look at everyone in the population, but that is usually practically impossible and an estimate based on a sample has to be used instead. Forensic scientists therefore have to do their best to obtain samples that are sufficiently representative of the relevant population that the output of the forensic analysis system is close enough to the true answer for that output to be useful. A way to assess how useful a forensic analysis system is, is to empirically test the performance of the system. This is known as empirical validation. It is done by presenting the system with a large number of test pairs that the tester knows to be same origin and a large number of test pairs that the tester knows to be different origin, and seeing how good on average the system's output is. Larger likelihood ratio values are better if the test pair is actually same origin and smaller likelihood ratio values are better if the test pair is actually different origin. For the test results to be meaningful, the test pairs must be sufficiently representative of the relevant population and sufficiently reflective of the conditions of the case under investigation. Requiring empirical validation was the number one recommendation in the 2016 PCAST report.

---

[4] President's Council of Advisors on Science and Technology (2016). *Forensic science in criminal courts: Ensuring scientific validity of feature-comparison methods*. https://obamawhitehouse.archives.gov/sites/default/files/microsites/ostp/PCAST/pcast_forensic_science_report_final.pdf

[5] Morrison G.S., Kaye D.H., Balding D.J., Taylor D., Dawid P., Aitken C.G.G., Gittelson S., Zadora G., Robertson B., Willis S.M., Pope S., Neil M., Martire K.A., Hepler A., Gill R.D., Jamieson A., de Zoete J., Ostrum R.B., Caliebe A. (2017). A comment on the PCAST report: Skip the "match"/"non-match" stage. *Forensic Science International*, 272, e7–e9. http://dx.doi.org/10.1016/j.forsciint.2016.10.018. Preprint available at https://www.newton.ac.uk/files/preprints/ni16050_0.pdf or https://ssrn.com/abstract=2860440

For real forensic data, as opposed to the simplified hair color example given above, sophisticated statistical models are used to calculate the probabilities that make up the likelihood ratio. There are many different statistical models and model variants that a forensic scientist can potentially choose from. The core of the journal article by Lund & Iyer focuses on variability in the likelihood ratio output due to choosing different statistical models or model variants. If a different statistical model or model variant is used, then the calculated value of the likelihood ratio will change, in the same way that if a different sample of the relevant population is used the calculated value of the likelihood ratio will change.

The topic of variability in likelihood ratio values in general is something I am very interested in. As Lund & Iyer mention in their article, in 2016–2017 I edited a special issue of the journal *Science & Justice*[6] in which contributors with differing views debated how to deal with variability in forensic likelihood ratio values. The special issue had an open call for position papers, and an open call for responses to those papers. Although Lund & Iyer's article was published in NIST's in-house journal, which does not accept papers or responses unless they are authored by NIST staff members, I see the core of their article as a valuable addition to that debate.

Although potentially useful for research on improving forensic analysis systems, the particular source of variability addressed in the core of the Lund & Iyer article is not actually of concern for a court of law. For a judge's decision on whether forensic testimony based on a particular forensic analysis system is admissible, whether some other system (or some other statistical model) works better is not a relevant consideration. What is relevant is whether the performance of that particular system has been empirically tested under conditions sufficiently similar to the conditions of the case at hand that the results of those tests will be relevant for the case at hand, and whether the results of those tests show the performance of the system to be good enough under those conditions. See Federal Rule of Evidence 702 and the Supreme Court rulings in *Daubert*, *Joiner*, and *Kumho Tire*. William C Thompson and I discussed this point in some length in an article published in the spring 2017 issue of the *Columbia Science and Technology Law Review*.[7]

Of course if forensic scientists called by prosecution and defense present very different results, then the court will want to enquire as to the reasons for the difference. Note, however, that such disagreements could occur irrespective of whether the likelihood ratio framework was used. If the forensic scientists'

---

[6] http://www.sciencedirect.com/science/journal/13550306/vsi?sdc=1

[7] Morrison G.S., Thompson W.C. (2017). Assessing the admissibility of a new generation of forensic voice comparison testimony. *Columbia Science and Technology Law Review*, 18, 326–434. http://www.stlr.org/cite.cgi?volume=18&article= morrisonThompson. Preprint available at https://www.newton.ac.uk/files/preprints/ni16053.pdf or https://ssrn.com/ abstract=2883767

conclusions were based directly on their untested subjective judgments, it would be very difficult to trace the causes for the disagreement or to choose between the different results. In contrast, transparent implementation of the likelihood ratio framework using relevant data and statistical models would allow the court to identify potential causes for major disagreement in results. And if no particular flaws were found, but both forensic analysis systems were empirically tested on test pairs that reflect the conditions of the case, then the jury could put more trust in the output of the system that was shown to have the best performance under those conditions. Transparent implementation of the likelihood ratio framework using relevant data and statistical models with empirical testing of performance under casework conditions is actually the solution to the problem.

The core of Lund & Iyer's journal article will only be accessible to readers with a sufficient background in statistics. The introduction, discussion, and summary sections may be more accessible. Those sections contain what is essentially a strawman argument, and present a conclusion that I consider to be unsupported by the material in the core of the article. The argument and conclusion are repeated in a more sensationalist manner in the press release.

The press release does not advocate that likelihood ratios not be used at all, but it leaves the strong impression that they should only be presented in court in exceptional circumstances. I and many others (e.g., the vast majority of presenters at the *International Conference on Forensic Inference and Statistics* held in Minneapolis in September 2017) [8] take exactly the opposite position: The likelihood ratio framework is the logically correct way for a forensic scientist to assess strength of evidence, and it would be an exceptional case in which it would not be appropriate for the forensic scientist to present a likelihood ratio in court. Further, I would argue that it is incumbent on the forensic scientist to carefully explain to the jury: (1) what they have adopted as the relevant population; (2) how they have obtained a sample that they believe is representative of that population and reflective of the conditions of the case, both for calculating the typicality probability and for empirically testing the performance of their forensic analysis system; and (3) what the results of the empirical testing are.

Lund & Iyer use what is essentially a strawman argument: They claim that likelihood ratio values are based on the forensic scientist's personal subjective beliefs, and that those may differ from what the jury's beliefs about the strength of evidence would be if the jury could analyze the evidence themselves (because, for example, the forensic scientist and jury had different initial assumptions). In such a situation it would not be appropriate for the jury to use the forensic scientist's likelihood ratio value as an indicator of the strength of the evidence. Unfortunately, there are many who take the extreme view that strength of evidence values can be assigned entirely on the basis of the forensic scientist's untested subjective judgment. They argue that such judgments can be made on the basis of the practitioner's training and experience. This is a view that was roundly condemned by the 2016 PCAST report. It is also a view which is probably much more prevalent among those who do not use the likelihood ratio framework than those who do. Although there are some advocates of the use of the likelihood ratio framework who also take the view that likelihood ratio values can be assigned entirely on the basis of the forensic scientist's untested subjective judgment (Lund & Iyer's strawman), it is not true that likelihood ratios must be based on untested subjective judgment. Many forensic statisticians argue for a rigorous calculation of likelihood ratios based on relevant data, quantitative measurements, and statistical models.

---

[8] http://www.cvent.com/events/icfis-2017-international-conference-on-forensic-inference-and-statistics/event-summary-6d357a9583224144866d64f44de367a2.aspx

There is no such thing as 100% objectivity, there will always be a degree of subjectivity, but the forensic analysis will be more transparent, more resistant to cognitive bias, more easily replicated, and more easily tested if subjective elements are restricted to decisions about what constitutes the relevant population and whether the data used to build and test the statistical models are sufficiently representative of the relevant population and reflective of the conditions of the case. The subjective judgments made by the forensic scientist on these matters must be carefully explained to the jury, and if the jury accepts that these judgments were reasonable and is satisfied that the empirically demonstrated level of performance is adequate, then the jury can rely on the value of the likelihood ratio output by the forensic analysis system. It is the empirically demonstrated degree of performance of the forensic analysis system that justifies the jury's use of the value of the likelihood ratio presented by the forensic scientist.

It could be that, based on the test results, the jury thinks the performance of the forensic analysis system is fair but not good. In such a case the jury might decide to use a more conservative value for the strength of evidence than the likelihood ratio value presented by the forensic scientist. Whether the jury uses the same strength of evidence as that presented by the forensic scientist is not, however, an issue specific to the presentation of likelihood ratio values. A forensic scientist could state unequivocally 100% that the defendant was the source of the evidence found at a crime scene (such a statement could not be justified either logically or via empirical testing, and statements of this type were condemned in the 2016 PCAST report), but the jury could decide that there is some probability that the forensic scientist is incorrect and thus decide to use a strength of evidence value of less than 100%. A forensic scientist who presents a likelihood ratio value and the results of empirical testing of the system that produced that likelihood ratio value, (1) presents something which is logically correct, and (2) gives the jury information that they could use to decide whether to accept that likelihood ratio value as-is, and, if not, that they could use to decide how much more conservative a value to use.

Some in the debate about how to deal with variability in likelihood ratio values advocate the use of statistical models which actually calculate more conservative likelihood ratio values when the output of the system would otherwise be overly variable. I have suggested that use of such models could be a practical solution acceptable to those with opposing philosophical views on the debate.[9]

Are there problems in calculating appropriate likelihood ratio values and in communicating their meaning to a jury? Yes, but the existence of these problems is not reason to reject the use of the likelihood ratio framework, it is reason to invest more in research aimed at solving these problems.

Much too often in current practice across many branches of forensic science, what is presented in court is logically untenable untested subjective opinion. A number of forensic scientists, forensic statisticians, legal scholars, and bodies such as PCAST are fighting for reform. Progress is slow, and is not helped by sensationalist presentations of spurious arguments and unjustified conclusions such as those in the NIST press release.

---

[9] Morrison, G.S. (2017). What should a forensic practitioner's likelihood ratio be? II. *Science & Justice*. http://dx.doi.org/10.1016/j.scijus.2017.08.004

Dr Geoffrey Stewart Morrison is a Canadian researcher who works in the fields of forensic speech science, and forensic inference and statistics. He is author of more than 40 academic publications in these fields. He is Reader in Forensic Speech Science, Centre for Forensic Linguistics, Aston University. He is also a member of Aston University's Systems Analytics Research Institute. His past appointments include: Adjunct Professor, Department of Linguistics, University of Alberta; Simons Foundation Visiting Fellow, Probability and Statistics in Forensic Science Program, Isaac Newton Institute for Mathematical Sciences; Scientific Counsel, Office of Legal Affairs, INTERPOL; and Director of the Forensic Voice Comparison Laboratory, School of Electrical Engineering & Telecommunications, University of New South Wales. He has been a Subject Editor and Guest Editor for the peer-reviewed journals *Speech Communication* and *Science & Justice*. He has collaborated on research and development projects with law enforcement agencies in Australia, Europe, and the United States, and has forensic casework experience in Australia, the United States, and the United Kingdom. More information about his work can be found at http://geoff-morrison.net/.